\documentclass[aps,pra,twocolumn,10pt,superscriptaddress]{revtex4-2}
\pdfoutput=1

\usepackage[final]{graphicx}
\usepackage{times,bbm,amsmath,amssymb}
\usepackage{epsfig,color}
\usepackage{xcolor}
\usepackage{hyperref}
\hypersetup{
    colorlinks = true
}
\usepackage{cleveref}

\usepackage[caption = false]{subfig}

\usepackage[greek,english]{babel}
\usepackage{thumbpdf,enumerate}
\usepackage{booktabs}
\usepackage{sidecap}
\usepackage[scaled=.8]{couriers}    
\usepackage{pstricks}
\usepackage{multirow}
\usepackage{placeins}
\usepackage{relsize}  
\usepackage{pst-grad,bm}
\usepackage{epigraph}
\usepackage{textcomp}
\usepackage{gensymb}
\usepackage{longtable}
\usepackage{booktabs}
\usepackage{gensymb}

\usepackage{soul}
\usepackage{ulem} 
\usepackage{acronym}

\usepackage{tikz}

\usepackage{bbold}
\usepackage{microtype}
\usepackage{listings}
\newcommand{\comment}[1]{}

\begin{document}

\title{Validation of a noisy Gaussian Boson Sampler via graph theory}

\author{Denis Stanev}
\affiliation{Gran Sasso Science Institute, Viale Francesco Crispi 7, I-67100 L’Aquila, Italy}
\author{Taira Giordani}
\affiliation{Dipartimento di Fisica, Sapienza Universit\`{a} di Roma, Piazzale Aldo Moro 5, I-00185 Roma, Italy}
\author{Nicol\`o Spagnolo}
\affiliation{Dipartimento di Fisica, Sapienza Universit\`{a} di Roma, Piazzale Aldo Moro 5, I-00185 Roma, Italy}
\email{nicolo.spagnolo@uniroma1.it}
\author{Fabio Sciarrino}
\affiliation{Dipartimento di Fisica, Sapienza Universit\`{a} di Roma, Piazzale Aldo Moro 5, I-00185 Roma, Italy}

\begin{abstract}
Quantum photonic processors are emerging as promising platforms to prove preliminary evidence of quantum computational advantage towards the realization of universal quantum computers. In the context of non-universal noisy intermediate quantum devices, photonic-based sampling machines solving the Gaussian Boson Sampling problem currently play a central role in the experimental demonstration of a quantum computational advantage. A relevant issue is the validation of the sampling process in the presence of experimental noise, such as photon losses, that could undermine the hardness of simulating the experiment. In this work, we test the capability of a validation protocol, that exploits the connection between Gaussian Boson Sampling and graphs perfect match counting, to perform such an assessment in a noisy scenario. In particular, we use as a testbench the recently developed machine Borealis, a large-scale sampling machine that has been made available online for external users, and address its operation in the presence of noise. The employed approach to validation is also shown to provide connections with the open question on the effective advantage in using noisy Gaussian Boson Sampling devices for graphs similarity and isomorphism problems, and thus provides an effective method for certification of quantum hardware.
\end{abstract}

\maketitle

\section{Introduction}

Quantum devices and quantum algorithms promise substantial advantages in many computational tasks. Such applications include as notable examples quantum computing, simulation, communication, and sensing. In recent years, an ever-increasing number of advancements is being made in this field, with the first experiments \cite{QuantumSupGoogle,QuantumSupChina,PhysRevLett.127.180501,PhysRevLett.127.180502,XanaduQAdv, Jiuzhang3} tackling the quantum advantage regime, namely the scenario where quantum devices are capable of outperforming classical computers in specific tasks. These results have thus opened the way for the development and application of noisy intermediate-scale quantum (NISQ) processors \cite{NISQ} for quantum-enhanced information processing. 

An example of a non-universal quantum processor is the Gaussian Boson Sampling (GBS) scheme. It is a variant of the original proposal of Boson Sampling (BS), which is a classically-hard computational problem that can be tackled through the use of dedicated quantum photonic devices. More precisely, GBS is the problem of generating samples from the photon-counting output distribution of indistinguishable Gaussian states of light after the evolution through a multi-mode random linear optical interferometer \cite{BosonSampling1, BosonSampling2, GBSHardness1, GBSHardness2, GBSHardness3}. This problem is intractable for a classical computer when the input states are indistinguishable sources of single-mode squeezed states. Then, a dedicated quantum photonic device can tackle such a task more efficiently, thus corresponding to a quantum computational advantage for a problem instance of sufficient size. The GBS problem has thus drawn attention in the photonic community due to the practical chance to achieve the quantum advantage regime with the technology available today. The latest GBS instances have reached the condition where the quantum device has solved the task faster than current state-of-the-art classical strategies in several experiments \cite{QuantumSupChina,PhysRevLett.127.180502, Jiuzhang3}, including the \emph{Borealis} machine by {the company} Xanadu \cite{XanaduQAdv}. Further interest in the GBS scheme is motivated by the connection between the sampling process and the problem of counting perfect matchings of arbitrary graphs. In fact, an important property of GBS is that it is possible to encode an adjacency matrix of a graph in the device by proper tuning of the interferometer parameters and the squeezing values. Thus, the collected samples could be used to learn several properties of the graph, and be employed to solve some relevant problems in the field such as finding the dense subgraphs and max-clique, simulating vibronic spectra, and graph similarity \cite{Bromley_2020}. Recently, GBS-based algorithms to solve such well-known problems in graph theory have been formulated \cite{GBSGraphTheory1, GBSGraphTheory2, GBSGraphTheory3}. First tests on quantum devices have been then performed on small-scale integrated photonics devices \cite{GBSGraphExperiment} and on larger dimensions in bulk optics through time- \cite{1LoopPaper,yu2023universal} and path-encoded interferometers \cite{Graphs_GBS_china}.

In recent years large efforts have been devoted to developing more efficient classical algorithms capable of simulating the sampling process \cite{clifford2017classical,Neville2017,Quesada_exact_simulation,Quesada_exact_simulation_speedup, oh2022classical,Bulmer_Markov_GBS, Drummond2022, popova2021cracking, cilluffo2023simulating, TensorNetworkGBS}, with the motivation of investigating the classical simulatability threshold of GBS devices. In this context, further studies had individuated sources of experimental noise that could undermine the hardness of the problem and allow for classical simulations \cite{Keshari_2016_conditions}, such as photon losses \cite{Oszmaniec_2018,GarciaPatron2019simulatingboson,Brod2020classicalsimulation, Qi_lossyGBS, ohlossy, liu2023complexity} and photon distinguishability \cite{Renema_2018_classical,Moylett_2019, Renema_partial_2020, Shi2022}. Thus, the benchmarking of a GBS experiment can be performed by following a validation approach, i.e. a test that discerns when samples are drawn from classical simulatable models. The methods vary from Bayesian tests \cite{GBSVal1, PhysRevLett.127.180502,martinezcifuentes2022classical}, statistical properties of two-point correlation functions \cite{GBSVal3,GBS_correlators, Giordani18_correlators} or higher order ones \cite{PhysRevLett.127.180502}, detector binning \cite{Bressanini2023}, properties of marginal probabilities \cite{Renema_partial_2020, renema2020marginal, villalonga2021efficient}, to a very recent method based on the feature vectors components of the graph encoded in the device \cite{TairaGBS}. 
 
In this work, we focus on testing graph theory based methods to assess the operation of a Gaussian Boson Sampler in the presence of experimental noise. As a testbench system for this analysis, we employed the Gaussian Boson Sampler Borealis \cite{XanaduQAdv}, which recently claimed quantum advantages and has been made available on the cloud on Xanadu Cloud \cite{Xanadu} and on Amazon Braket \cite{Braket}. Some properties of the device have been recently investigated in Ref. \cite{BorealisValQCS} by remote users. The authors measured the quadrature coherence scale to find genuine signatures of the features of single-mode bosonic systems in the phase space representation. Here, we analyze the capabilities of the method based on graphs feature vectors \cite{TairaGBS} to exclude that Borealis samples collected in a real experiment subjected to noise can be compatible with some relevant classically-simulatable modes, that is, thermal, coherent, squashed or distinguishable particles samplers. First, in Section \ref{sec:Background} we will provide background information on Gaussian Boson Sampling. Then, in Section \ref{sec:Structure} we will review the main features of the structure of Borealis. After that, in Section \ref{sec:Interfacing} we will explain the process for interfacing with Borealis, while finally in Section \ref{sec:Results} we will analyze the collected data and perform the validation of Borealis against the aforementioned alternative models.

\section{Background on Gaussian Boson Sampling and connection with graph theory}
\label{sec:Background}

GBS is a linear-optics scheme to generate samples drawn from the photon-counting distribution generated by Gaussian light sources at the outputs of a multi-mode interferometer \cite{GBSHardness1,GBSHardness2, GBSHardness3}. Such a sampling task is hard to simulate for certain classes of Gaussian states such as indistinguishable sources of single-mode squeezed vacuum (SMSV) states. The hardness is preserved in noisy experimental conditions as long as the levels of losses and photon distinguishability are limited. 

Consider $m$ independent sources of Gaussian states $\rho_i$ without displacement such that the input is $\rho = \bigotimes_{i=1}^m \rho_i $ with a $2m \times 2m$ covariance matrix $\sigma$. The probability distribution of detecting $n$ photons in the configuration $\Vec{n}=(n_1, n_2,..., n_m)$, where $n_i$ is the number of photon in the output $i$ and $\sum_{i=1}^m n_i=n$, is
\begin{equation}
    P(\Vec{n}) = |\sigma_Q|^{-1/2} \frac{\text{Haf} (A_{\Vec{n}})}{\prod_{i=1}^m n_i!}.
\end{equation}
The quantity $\sigma_Q$ corresponds to $\sigma + I_{2m}$, being $I_{2m}$ the $2m \times 2m$ identity matrix, and Haf to the hafnian operation of the $2n \times 2n$ sub-matrix $A_{\Vec{n}}$. Such a sub-matrix ($A_{\Vec{n}}$) is individuated by taking $n_i$ times the $i$th row and the $i + m$-th column of $A$, and the hafnian is the operation that counts the number of perfect-matching in a graph represented by a symmetric adjacency matrix \cite{Caianiello1953}. The whole $2m \times 2m$ matrix $A$ has the following structure
\begin{equation}
    A = \left (\begin{array}{cc}
         B&  C \\
          C^T& B^* 
    \end{array}
    \right ),
\end{equation}
where $B$ is an $m \times m$ symmetric matrix and $C$ an $m \times m$ hermitian one. Both blocks depend on the transformation $U$ implemented by the interferometer and on the input Gaussian state. For example, the matrix representing $m$ indistinguishable SMSV states with squeezing parameters $s_i$ in a lossless experiment has $C=0$ and $B = U \text{diag}(\tanh {s_1}, \dots, \tanh{s_m}) U^T$. In this case, the output probabilities can be evaluated according to the following expression:
\begin{equation}
    P(\Vec{n})_{\mathrm{SMSV}} = |\sigma_Q|^{-1/2} \frac{\vert \text{Haf} (B_{\Vec{n}}) \vert^2}{\prod_{i=1}^m n_i!}.
\end{equation}
In this scenario, the adjacency matrix of a graph can be encoded in $B$, through the Takagi-Autonne factorization, by tuning $U$ and $s_i$ values. This provides a direct link between the output of specific instances of Gaussian Boson Sampling and graph theory. For this reason, GBS with SMSV states had attracted lots of attention not only for proving a quantum advantage in the sampling task but also for applications on graphs. Indeed, GBS devices have been suggested as possible tools to tackle problems such as finding densest sub-graphs, the max-clique \cite{GBSGraphTheory1,1LoopPaper} and graphs similarity \cite{GBSGraphTheory2, GBSGraphTheory3, GBSGraphExperiment}. The opposite scenario with respect to pure SMSV inputs is a thermal sampler, in which there is no coherence among the possible number of emitted photons that result in $A$ matrix with $C\neq 0$ only. Note that the latter case correspond to a  classically simulatable model, given that an efficient sampling algorithm from thermal inputs can be defined.

Many approaches have been proposed to validate GBS experiments in the quantum advantage regime \cite{GBSVal1, GBSVal3, GBS_correlators, Giordani18_correlators,TairaGBS}. The goal of a validation algorithm is to exclude that the samples are drawn from classical simulatable distributions, such as the outputs of thermal, coherent, squashed and distinguishable Gaussian samplers. In this work, we apply a recent method introduced in Ref. \cite{TairaGBS} that makes use of graphs feature vectors estimated from GBS samples. The components of such vectors are called \emph{orbits}, which result from a coarse-graining of the output configurations. More precisely, the method consists in the classification of different samplers in the space spanned by the three feature vector components identified by the probability of the orbits $O_1 = [1,1,1,1,...]$, $O_2=[2,1,1,1,...]$ and $O_3=[2,2,1,1,...]$ for a given number $n$ of detected photons. An orbit $\vec{n}$ is defined as the set of possible index permutation of $\vec{n}$. The three orbits $\{O_1, O_2, O_3\}$, employed for the validation method, collect respectively the output states in which the number of photons in the modes is 0 or 1, in which only one detector measures two photons, and in which two modes host two photons. Plotting the feature vectors in the space $\{O_1, O_2, O_3\}$, retrieved by summing the normalized frequencies of the output configuration corresponding to each orbit, can be used to distinguish between various types of classical samplers Indeed, the orbit distributions for the different models are characterized by different behaviours, such as for instance lying on different hyperplanes. We underline that such a choice is effective in the $n \ll m$ regime, in which we expect that bunching configuration have a very small probability to occur. In general, one has to chose the orbits that have the highest probability in order to be estimated from a finite sample with a good accuracy.

\section{Structure of Borealis as a time-bin GBS}
\label{sec:Structure}

\begin{figure*}[ht!]
    \includegraphics[width=1\textwidth]{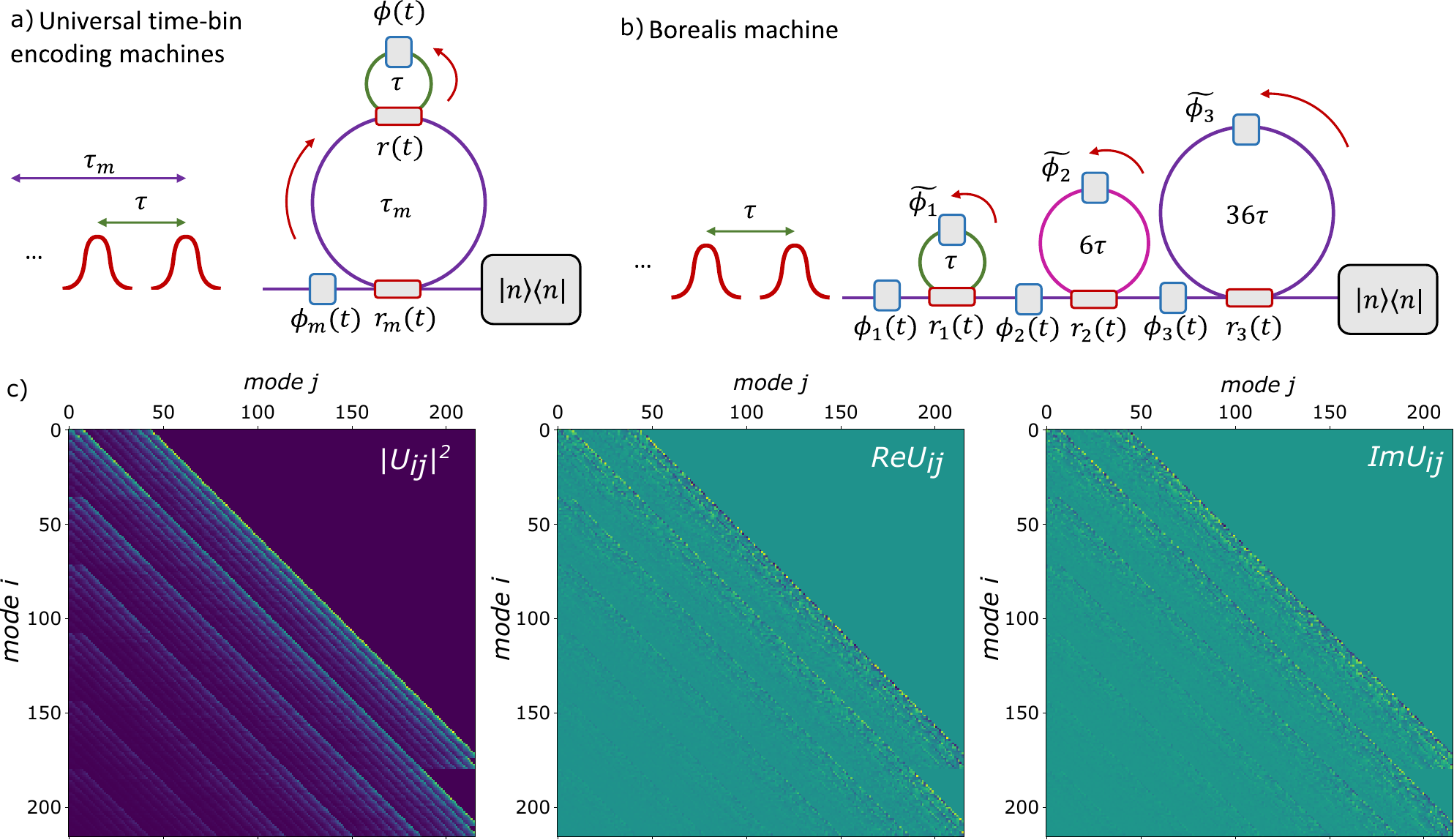}
    \caption{\textbf{Time-bin encoding machines and Borealis structure.} a) Example of universal time-bin interferometers. In such an encoding the optical modes are discrete time bins. The unitary operation over the modes is performed by two concatenated fiber loops. The shortest loop covers the time separation between two consecutive time bins, while the longest one covers the whole time duration of the number of modes. The same squeezing source is excited at each pulse, and photon counting measurements are performed at the end of the evolution. b) Structure of Borealis \cite{BorealisArchitecture, XanaduQAdv}. The interferometer comprises three consecutive loops of increasing length. c) Example of a unitary matrix $U_{ij}$  performed by Borealis that shows the limited connectivity among the modes due to the loops structure. }
    \label{fig:BorealisConnectivity}
\end{figure*}

Borealis\cite{XanaduQAdv} is based on time-domain multiplexing (TDM) \cite{BS_loop_architcture,BS_China_loop,TDM} with limited connectivity. A single squeezed-light source emits batches of $m$  time-ordered squeezed-light pulses that interfere with one another with the help of optical delay loops, programmable beamsplitters (BSgate), and phase shifters (Rgate). The loops arrangement reported in Fig. \ref{fig:BorealisConnectivity}a is an example of a universal time-bin interferometer able to encode any operation over the modes (see also Refs. \cite{BS_loop_architcture,BS_China_loop}). The transformation is controlled by time modulations of the splitting ratios of two BSgates [$r_m(t)$ and $r(t)$] and of the two phases of the Rgates [$\phi_m(t)$ and $\phi(t)$]. The two loops are concatenated and cover the time separation $\tau$ between consecutive bins and the whole time duration of the $m$ modes, $\tau_m = m  \tau$, respectively.  This architecture has been recently employed to realize a programmable and universal interferometer encoded in the time-bin for GBS applications \cite{yu2023universal}.

The loops structure of Borealis does not follow the universal layout. Fig. \ref{fig:BorealisConnectivity}b shows the time-bin interferometer that comprises 3 consecutive loops with 3 tunable BSgates $r_{1,2,3}(t)$ and likewise tunable Rgates $\phi_{1,2,3}(t)$. The tunable phase-shifters are limited to the range $[-\frac{\pi}{2}, \frac{\pi}{2}]$. There are further 3 static phases $\tilde{\phi}_{1,2,3}$ that cannot be controlled by the user and represent the optical phases of each loop. The time separation between the bins is $\tau = 167$ ns. Then, the time that covers the evolution of the 216 logical modes plus the 43 ancillary modes of the device is 43.5 $\mu$s and thus coincides with the time needed to obtain one sample from the device. The length of the 3 loops covers a time delay equal to $\tau$, $6\tau$ and $36\tau$ respectively. The squeezer has 4 settings for the squeezing parameter, ``low'', ``medium'', ``high'' and 0. The squeezing must be set to one of the allowed values and it cannot be modulated over the temporal modes. The setup ends with the single photons detection stage. The time bins are translated into 16 different spatial modes by a time-to-space demultiplexer system. Single photons are then detected by 16 photon-number-resolving detectors \cite{XanaduQAdv}.

The design of Borealis has been chosen to be a compromise between the need for a programmable device and the requirement to achieve the quantum advantage regime. The architecture does not allow for full connectivity between the modes, with a layout that has been recently shown to reduce the required dimension for simulation via tensor networks \cite{TensorNetworkGBS}, while still permitting to cover high-dimensional spaces with minimized losses and optical resources \cite{GBSHardness3}. For connectivity, we mean the number of modes with which each of the modes interacts directly. In Borealis, for example, the loop structure allows each mode to be connected with at most 6 time bins. Thus, the 3-loop connectivity of Borealis puts some restrictions on the matrices that can be represented, as we can see in Fig. \ref{fig:BorealisConnectivity}c. The transfer matrix is limited to being 0 on the upper half of the diagonal. The moduli of the matrix elements decrease when descending from the diagonal. They increase only every 6 and 36 modes, but are still smaller than the values at 6 or 36 modes above. Other restrictions derive from the ranges of tunability of the Rgates that do not allow the encoding of meaningful graphs adjacency matrices inside the device. 

\begin{figure*}
    \centering
    \includegraphics[width=\textwidth]{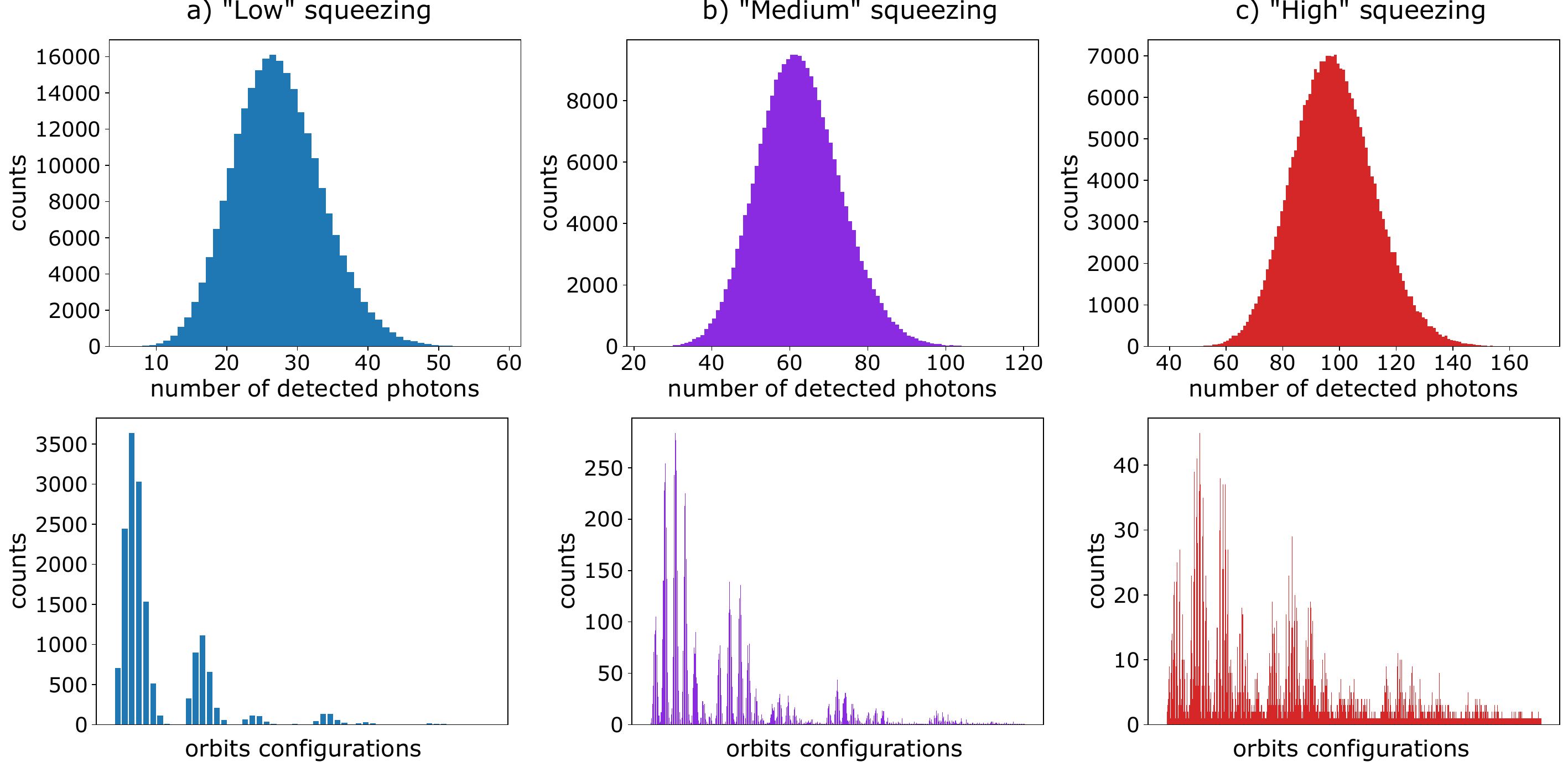}
\caption{\textbf{Detected photons and orbits probability distributions.} On the top row, the distributions of the number of detected photons for the three levels of squeezing, a) ``low'', b) ``medium'', c) ``high''. On the bottom, the distributions of the orbit configurations for the number of detected photons associated to the highest probability, namely $n=$26 for ``low'' squeezing, $n=$62 for ``medium'', and $n=$97 for ``high'' squeezing. The x-axis represents the various orbits configuration ordered as follows, first $[1,1,1,\dots]$, then $[2,1,1, \dots]$, $[2,2,1,\dots]$, etc. The y-axis represents the number of samples for that orbit configuration. The size of each sample was 250000 shots.}
        \label{fig:Squeezings}
\end{figure*}

\section{Interfacing with Borealis}
\label{sec:Interfacing}

Borealis was accessed through Amazon Braket \cite{Braket}, and the parameters sent were the squeezing values, the parameters of the 3 beamsplitters and the parameters of the 3 phase shifters for each time mode, as well as the number of shots (i.e. the number of samples). The parameters sent are for 259 time modes, even though the number of outputs that we show is 216. This is because the first 43 modes are used to fill the loops, and thus the output for those first 43 modes is basically background noise. The device thus has 216 ``logical'' modes, while it has 259 ``physical'' modes. 

When interfacing with Borealis one also downloads the device certificate, which describes the current calibration of the device. It contains information such as the loop phases, the squeezing values for the various settings, and the various efficiencies. The certificate distinguishes between 3 types of efficiencies: 
\begin{itemize}
    \item common efficiency, which corresponds to the balanced losses of the device independent from the implemented circuit.
    \item loop efficiencies, which correspond to the losses of each loop (3 values, one for each loop).
    \item relative channel efficiencies, which correspond to the relative efficiencies of the detectors (16 values, one for each detector).
\end{itemize}
The amount of the losses changed on a daily basis. The average number of photons in the output from Borealis was usually around $\sim$75\% less than that of a lossless simulation with the same parameters. This is due to the slightly different operating conditions of the cloud version of Borealis with respect to the specifications reached in Ref.\cite{XanaduQAdv}.

The information included in the device certificate is necessary to perform numerical simulations with alternative models such as thermal samplers, which are used for comparison to assess the operation of Borealis. One aspect to be taken into account regarding the device certificate is that it is measured once a day, right before the 2-hour period in which Borealis is available. Thus, it could be less representative of the state of Borealis in those runs far away from the calibration. Indeed, the values of these efficiencies seem to oscillate over time, as we will show in Section \ref{sec:Results}. After some discussion with Xanadu's team, it turned out that these oscillations in the common efficiency are mainly due to variations in temperature. The scale of these variations in efficiency varies on a daily basis, with some days being more stable than others. 

Another aspect to be taken into account regarding the device certificate is that simulations based on the included parameters, in particular with respect to noise, do not always accurately reproduce some features of the machine output. For instance, in some cases we found that the difference in the average number of detected photons per shot between the simulations performed according to the certificate parameters and the measured samples was small.  This indicated that the loss estimation in the device certificate was accurate. In other cases, the mismatch between simulation and experiment was significantly higher, around 1.5-2 photon difference on an average number of photons of $\sim 26$, a value not compatible with statistical fluctuations. In these trials, the device certificate did not represent the physical device with sufficient accuracy within the statistical errors of the apparatus. This issue started to be more evident after one maintenance period that was made near the end of January 2023. 
Such discrepancy is relevant when trying to compare the experimental samples with simulations of classical samplers in the validation stage of the device, as it is no longer possible to have a truly accurate simulation for those days. 

\section{Borealis data acquisition and validation}
\label{sec:Results}

We will now analyze the results of our runs on Borealis, and compare them with some simulations. All of the code to interface with Borealis, perform the simulations, and analyze the data was written in Python 3 \cite{Python3}, and was based mainly on the libraries Strawberry Fields \cite{Strawberryfields} and The Walrus \cite{TheWalrus}. 

In Fig. \ref{fig:Squeezings} we report the distribution of the number of detected photons at the Borealis outputs for the three squeezing levels larger than zero. Our goal is to apply the orbits method to validate our data against classically simulatable models such as thermal, coherent, squashed and distinguishable SMSV input states. The method of comparing the orbits is effective even in the quantum advantage regime as long as the number of modes $m$ is considerably larger than the number of detected photons \cite{TairaGBS}. The reason is that in such a condition the probability distribution of the various orbits configuration is peaked on a few orbits. This is evident in Fig. \ref{fig:Squeezings} in which the experimental orbits distributions for the three squeezing levels are reported. For this reason, all the runs shown in this section were made with the squeezing parameter set to ``low'', which satisfies the condition $m \gg n$. The results of Fig. \ref{fig:Squeezings} also motivate the performed choice on the orbits for the application of the validation technique, given that they corresponds to highly populated ones.
 
\begin{figure}[h!]
    \centering
    \includegraphics[width=0.6\linewidth]{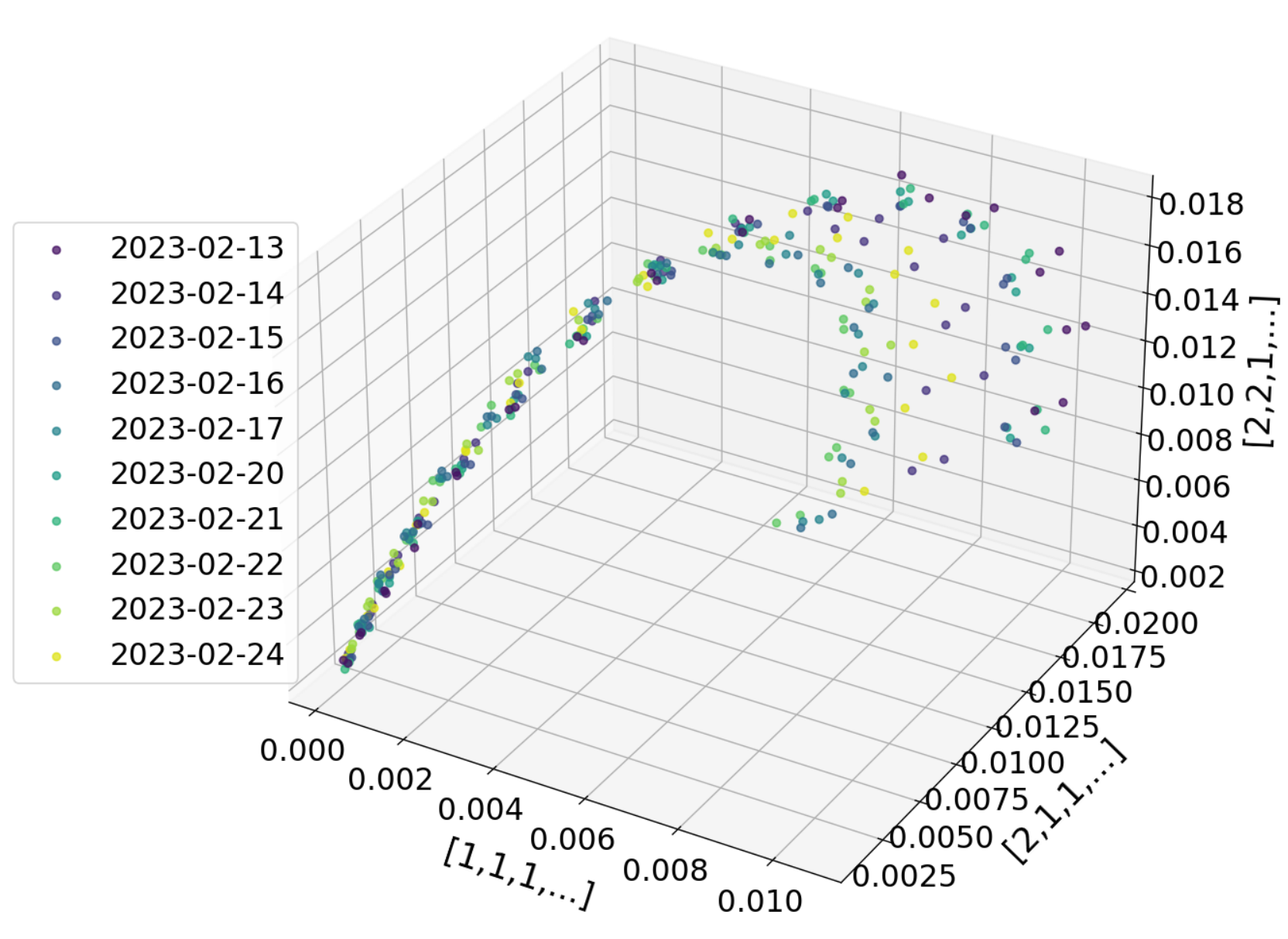}
    \caption{\textbf{Stability check of Borealis total efficiency over days.} Orbits distribution for the 2 weeks of runs with the same circuit settings. The variations in the different days are due to changes in the total efficiency of the apparatus, which is mostly dominated by the variations in common efficiency. The points correspond to the three orbits' probability for detected photons in the range between 18 and 32 photons. More precisely, each point on the plot is associated with a specific detected photon number. The points at the bottom left corner correspond to 32 photons, while the points on the right are the orbits for 18 post-selected photons.}
    \label{fig:2Weeks}
\end{figure}

To check the stability of the results over time, we ran sampling experiments with the same settings of the circuit for two weeks. We executed two runs of 250000 shots each day from Monday to Friday for these two weeks. The parameters we used are the following: all beamsplitters are set to a transmissivity of $50 \% $, and all phase shifters are set to $0$. We notice that the observed changes in the average number of detected photons in the outputs during the days impacts significantly the probability of orbits $\{O_1, O_2, O_3\}$. This variation is likely mostly due to changes in the common efficiency between days, as mentioned in the previous section. The plot of the three orbits probability for the 2 weeks is shown in Fig. \ref{fig:2Weeks}, while in Table \ref{tab:2Weeks} we report the average number of photons per each run of the day during the two weeks.

\begin{table}[h!]
\begin{center}
\begin{tabular}{ c|c c c c c }
\toprule
 Day & 13/02 & 14/02 & 15/02 & 16/02 & 17/02 \\ 
 \hline
 Run 1 & 23.86 & 24.98 & 24.20 & 25.77 & 26.03 \\  
 Run 2 & 23.61 & 24.64 & 24.21 & 26.01 & 25.81    \\
 \toprule
 Day & 20/02 & 21/02 & 22/02 & 23/02 & 24/02 \\ 
 \hline
 Run 1 & 24.15 & 23.85 & 25.98 & 25.45 & 25.21 \\  
 Run 2 & 24.16 & 23.99 & 26.05 & 25.43 & 24.70  \\
 \bottomrule
\end{tabular}
\end{center}
\caption{\textbf{Variation in the number of detected photons over days.} Average number of photons in each run of the 2 weeks acquisitions performed by setting the same matrix. The relative standard deviation caused by the limited number of shots (250000) is $\sim$ 0.04\% , which leads to a standard deviation in the mean photon number of around 0.01 photons. We thus observe that the mean number of photons in Borealis varies on each run performed in different days, and in some days this can even vary from a run to the next one, resulting in the orbits for the two runs of that day being further apart than the expectations.}
\label{tab:2Weeks}
\end{table}

\begin{figure*}
    \centering
    \includegraphics[width=\textwidth]{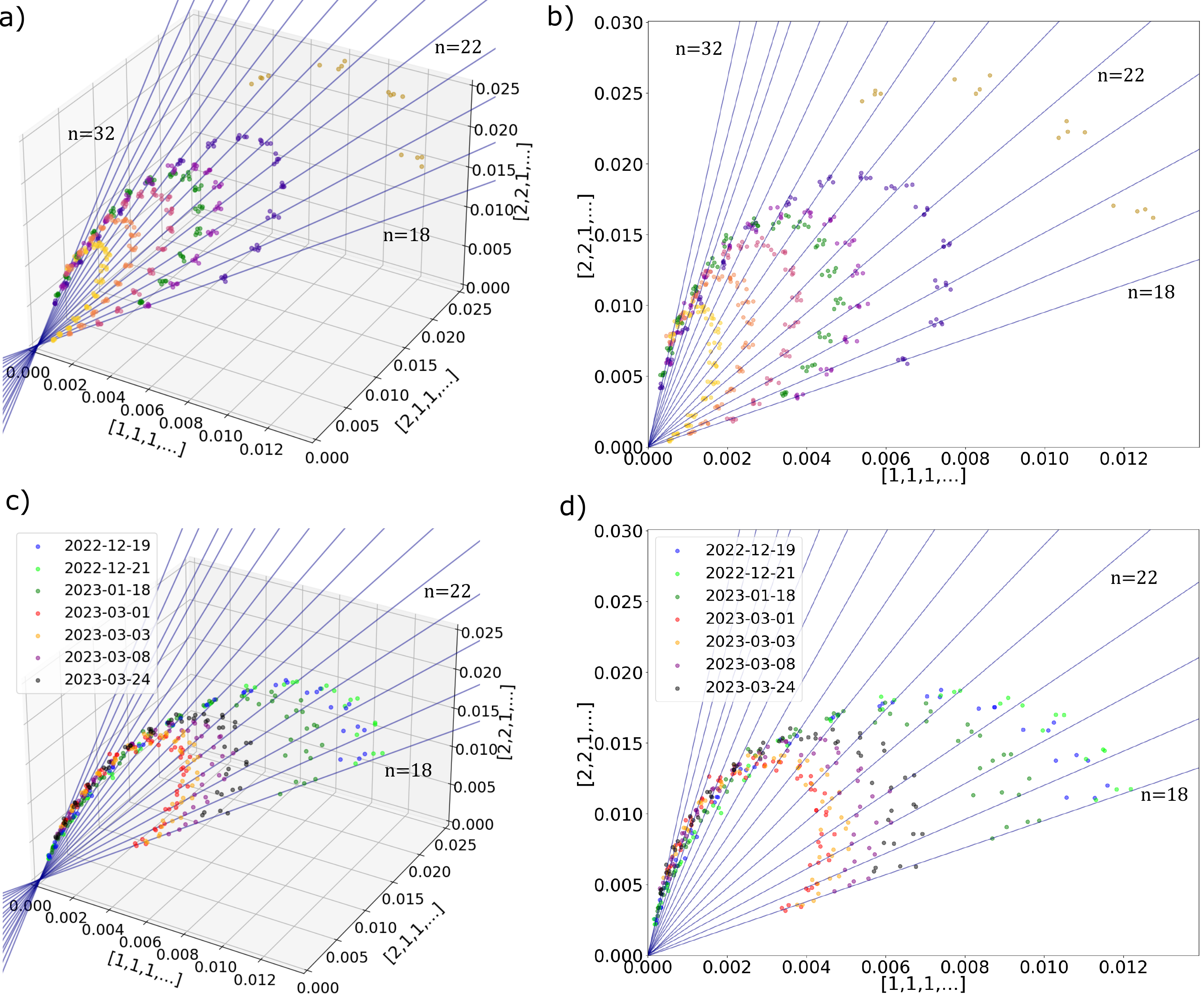}
    \caption{\textbf{Role of losses in the orbits estimation.} a-b) Orbits for the thermal sampler simulations. All points, except the green and dark gold ones, represent orbits where only the common efficiency has been changed. The lower the efficiency, the larger the radius of the orbits. The green points represent the case where one of the detectors was turned off, thus producing an unbalanced loss. The dark gold points are from the lossless indistinguishable SMSV states simulation. The details on the parameters used for the simulations can be found in the Supplemental Materials. The lines are obtained by making a linear fit on all points with the same number of detected photons, excluding the points of the orbit with unbalanced losses and the SMSV simulation. Each line represents a different number of photons. We highlighted the number of corresponding photons for some of the lines in the figure. The effect of balanced losses is to move points along the lines. The unbalanced ones move the points away from the lines, but keep them on the same hyperplane. However, such an effect is evident only in the case of a strong amount of unbalancement. The lossless SMSV states points are not on those lines, but still on the same hyperplane. We report the projection of the orbits in the plane in b). c-d) Experimental orbits from the runs on Borealis. The lines are the same as the panel above calculated from the simulation with thermal states.}
    \label{fig:OrbitsThermal}
\end{figure*}

\begin{figure*}[t]
    \centering
    \includegraphics[width=1\textwidth]{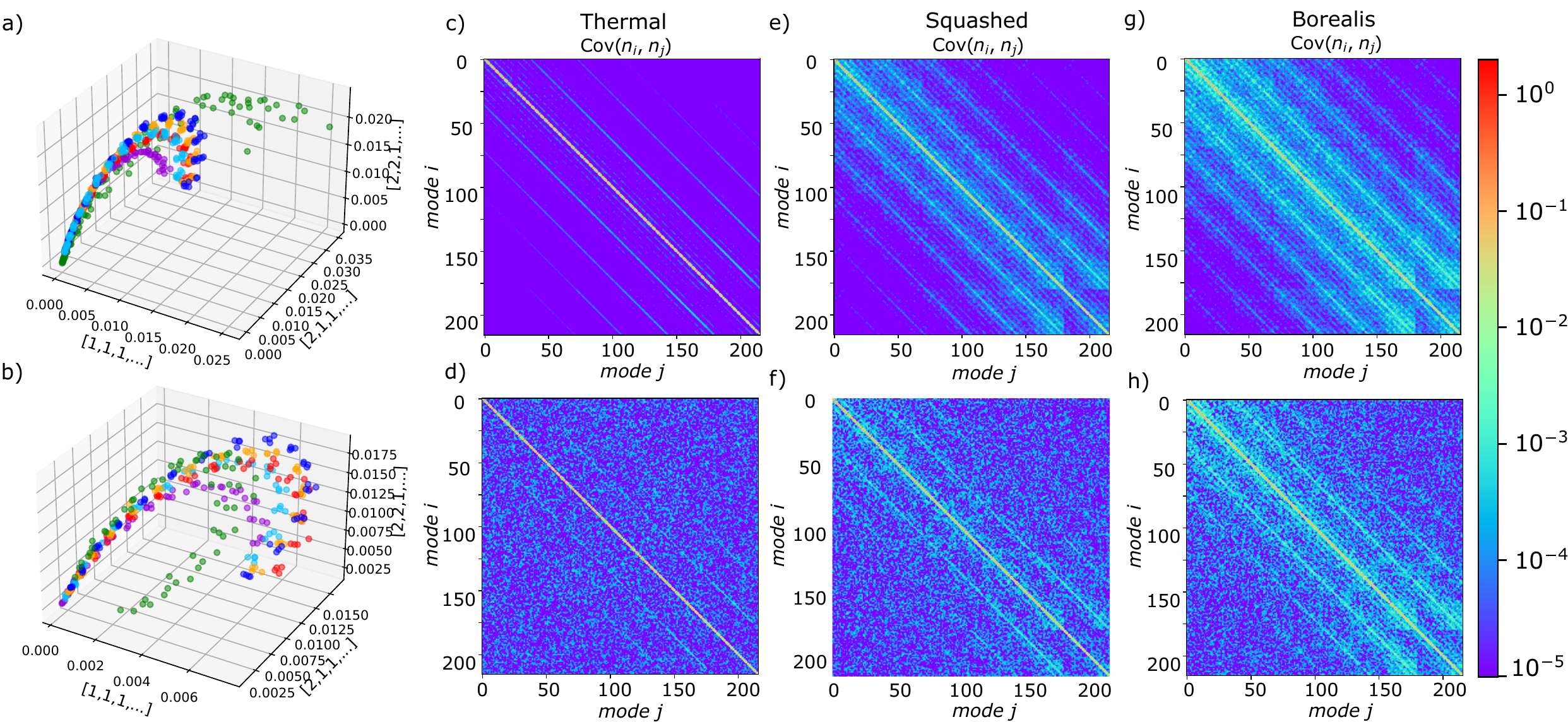}
    \caption{\textbf{Validation of Borealis samples.} a-b) Comparison of the orbits of Borealis (red), thermal states simulation (blue), squashed states simulation (orange), second-order greedy sampler with the ideal $U$ transformation and no loss (purple) and with the effective transfer matrix and loss (cyan), and coherent states simulation (green). The coherent, squashed, lossy greedy and thermal states simulations used the same parameters and device certificate of the Borealis runs. All the points were calculated on a sample size of 250000 shots with squeezing set to ``low''. a) Borealis data collected in different days with different $U$ transformations. Such runs display a similar average number of detected photons and device certificates. b) Borealis data collected on the same day with different unitary transformations. In that day, the fluctuation in the average number of detected photons between runs on Borealis was compatible with statistical fluctuations, and the deviations of the average number of detected photons predicted by the simulations according to the certificate with those measured on the device were small. c-h) Comparison of the covariances of lossy thermal states, lossy squashed states, and Borealis for a given transformation $U$ of the device. The same circuit parameters were used for the simulations and the measurements with Borealis. The same device certificate was used for all the simulations. c) Covariance of the simulated thermal sampler with losses.  d) Covariance of the thermal sampler with losses calculated from 250000 samples, to reproduce the additional noise due to a limited number of samples.  e) Simulated covariance of a squashed light sampler with the losses of the Borealis circuit. f) Covariance of squashed light sampler from a finite sample of 250000 shots. g) Simulated covariance of an SMSV GBS with losses. h) Covariance of Borealis samples.}
    \label{fig:Coherent}
\end{figure*}

We now discuss the validation of the samples collected from Borealis. Based on the results reported in Ref.\cite{TairaGBS}, we expect that indistinguishable SMSV GBS and indistinguishable thermal samples display orbits on the same hyperplane in the space spanned by $\{O_1, O_2, O_3\}$. Furthermore, in the lossless case thermal and squeezed light GBS have a similar spread of points in the orbits, while coherent has a spread of points that is much larger (see also Supplemental Materials). The presence of photon distinguishability results in a change of such a hyperplane for both squeezed, thermal and coherent light. The effect of the photon losses is to change the GBS orbits towards the thermal sampler \cite{TairaGBS}. In fact, both balanced (the common efficiency in Borealis) and unbalanced (loop and relative detector efficiencies) losses lead to changes in the orbits but always on the same hyperplane. Indeed, introducing losses to a Gaussian state corresponds to mixing it with the vacuum, thus increasing its thermal component. Hence, this effect does not alter the hyperplane of the orbits. Furthermore, in the presence of unbalanced transmissions between the channels, this introduces also changes to the transfer matrix of the device, which does not move the orbits in different hyperplanes. In Fig. \ref{fig:OrbitsThermal}a and b, we show in more detail such behaviors. The simulation considers thermal light inputs evolving through the transformation performed by the Borealis circuit with different levels of common efficiencies. We set the parameters of the circuit as follows: the transmittance of BSgates was set randomly between 0.4 and 0.6, while the phase shifters were set to random values in their full range $[-\frac{\pi}{2}, \frac{\pi}{2}]$. The points that correspond to the same number of detected photons move along lines. More precisely, the orbits somehow expand when the common efficiency decreases, and shrink when the efficiency increases. Unbalanced losses, on the other hand, tend to alter the orbits, moving the points away from those lines, but still keeping the orbit on that same hyperplane. Furthermore, from this analysis other imperfections, such as small errors in the squeezing parameters, beam-splitter ratios or interferometer phase shifts, are expected to provide only second order changes in the orbits. Errors in the squeezing parameter does not alter significantly the class of gaussian state, and thus the orbits would move on specific lines of the same hyperplane. Additionally, errors in beam-splitter ratios or phase shifts would correspond to slight changes in the transfer matrix, which do not alter significantly the orbits. We also made an approximate simulation of the orbits of an ideal GBS, as can be seen in Fig. \ref{fig:OrbitsThermal}a-b. From this simulation we observe that those orbits reside on the same hyperplane of the thermal ones, as expected. However, they do not align with the lines of the thermal sampler.

In Fig. \ref{fig:OrbitsThermal}c-d, we compare runs taken in around 3 months, from the end of December 2022 to March 2023, with a break in January for maintenance. The circuit settings were the same that we employed for the numerical simulations with thermal light. The changes in the orbits are compatible with changes in the common efficiency observed during the time of data collection. The other interesting behavior is that the orbits seem to follow closely the lines from the thermal simulations. However, they do to not match exactly those lines. Thus, according to this tests Borealis appears to behave as a lossy GBS with squeezed states, which suggests that the orbits may be intermediate between a GBS with indistinguishable thermal light and a lossless squeezed light GBS. Furthermore, the orbits still lay on the same hyperplane of the thermal, as for the case of a lossy GBS with indistinguishable emitters. This allows us to exclude that the samples derive from a distinguishable particle sampler. Since we have shown that changes in losses do not change the hyperplane of the thermal simulation, from our analysis we can restrict the hypothesis to a scenario where Borealis is either a thermal with losses, a coherent with losses, a squashed with losses, or an SMSV GBS with losses. As a further test, in the following we will also consider an additional mockup corresponding the greedy sampler \cite{villalonga2021efficient,XanaduQAdv}, that mimics samples with the correct marginals up to a chosen order $k$. More specifically, to further highlight the relevance of losses for Borealis, we considered both samples generate via the greedy algorithm for a lossless device implementing the ideal unitary $U$, with squeezing parameters adapted to lead to the correct average photon number, and a lossy case where the parameters of Borealis device certificate have been used to generate the greedy samples. 
\begin{figure*}[ht!]
    \centering
    \includegraphics[width=1\textwidth]{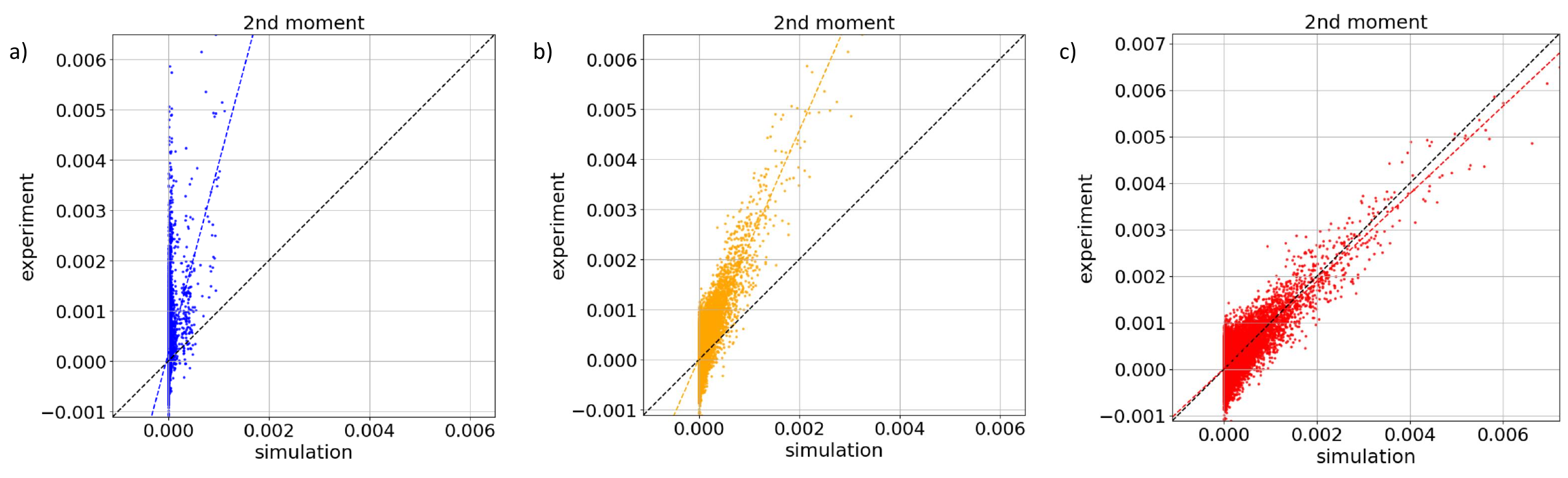}
    \caption{\textbf{Two-point correlators.} Scatter plot of two-point correlators $C_{ij}$ of samples collected from Borealis (experiment) as a function of the corresponding values calculated with different possible models for the device (simulation). a) Comparison with a thermal sampler. b) Comparison with a squashed state sampler. c) Comparison with a lossy single-mode squeezed vacuum states GBS. Black dashed line: the expected trend for the ground truth, corresponding to a linear function with a slope equal to 1. Colored dashed lines: trends obtained by a linear fit of the data shown in each figure.}
    \label{fig:two-point}
\end{figure*}
As explained in Ref.\cite{XanaduQAdv}, we considered as a relevant case for Borealis the scenario with marginals order $k=2$. We will now proceed to show that it is not a coherent light sampler, and then we will show that it is closer to an SMSV GBS than a thermal or a squashed one and to analyze the greedy sampler scenario. 

In Fig. \ref{fig:Coherent}a we compare data from runs of 250000 samples from Borealis on different days but with the same average number of detected photons (red points), with simulated samples from coherent (green), squashed (orange), lossless greedy (purple), lossy greedy (cyan) and thermal light (red). In panel b we consider runs performed on the same day in which the device certificate was providing an accurate estimate of the noise in the experiment due to photon losses. The orbits of coherent light are distant from Borealis data in both cases, and have a significantly bigger spread than the orbits of Borealis and of the thermal simulations. As a side note, we can also see that the orbits of the coherent sampler are distinct from those of a thermal sampler. In fact, even the mean photon number is different between the simulations of coherent and thermal samplers, indicating that losses affect the two class of states differently, given their different photon statistic. On the other hand, both Borealis and the thermal sampler seem to be affected in a very similar way by the losses, at least in terms of the mean photon number. We also observe that the samples from Borealis can be distinguished from the lossless greedy sampler, while the orbits in the lossy case are closer to the one of Borealis, and thus the discrimination of this mockup needs to be supported by other methods as those used in \cite{XanaduQAdv}. This result further highlights the relevance of losses in the analysis of the samples from  Borealis. Now, we are left only with the task of showing whether Borealis is closer to a thermal sampler, a squashed sampler, or an SMSV GBS with losses. Squashed states are the closest classical model to the lossy SMSV sampler. Samples drawn from the probability distribution of squashed states are obtained using classical mixtures of coherent states \cite{martinezcifuentes2022classical, Jahangiri_point_process}, similarly to the thermal sampler \cite{wcqoscct}. 

As we can see in Fig. \ref{fig:Coherent}a-b, the orbits of Borealis and of the thermal and squashed simulations are similar. In particular, the distances between the orbits of the two classical models corresponding to a given number of detected photons are compatible with the ones from Borealis within the typical orbits dispersion due to the intrinsic fluctuactions of Borealis parameters in time. This means that in this case, the orbits alone are not enough to ascertain whether Borealis is more likely to be a GBS with SMSV input states, a squashed sampler, or a thermal sampler. To further distinguish between these three cases, we look also at their respective sample covariance $\mathrm{Cov}(n_i,n_j)$. These quantities form the matrix of the two-point correlators in the output modes $i,j$, with matrix elements $C_{ij}$ defined as $C_{ij} = \mathrm{Cov}(n_i, n_j) = \langle n_i n_j \rangle -\langle n_i\rangle \langle n_j \rangle$. These quantities are informative to help in discerning the nature of the sampler as shown in the theoretical works \cite{GBSVal3, GBS_correlators} and in previous experiments \cite{Giordani18_correlators,QuantumSupChina,PhysRevLett.127.180501,PhysRevLett.127.180502,Jiuzhang3,XanaduQAdv}. Fig. \ref{fig:Coherent}c-h reports the simulated covariance matrices and the one calculated from a finite sample of 250000 shots. From this analysis, the behavior of Borealis is closer to the one of a SMSV GBS with losses, rather than a lossy thermal sampler or a lossy squashed sampler. This conclusion is supported by performing a more quantitative analysis of the two-point correlators. In Fig. \ref{fig:two-point} we compare the values $C_{ij}$ of samples collected from Borealis with the expected values calculated with the different possible models. We observe in Fig. \ref{fig:two-point}a-b, where the data are compared with a thermal and a squashed sampler, that these models do not provide an accurate explanation of the collected samples, as confirmed by the slopes of a linear fit corresponding respectively to $\sim 3.76$ (thermal) and $\sim 2.28$ (squashed). Conversely, the comparison of the correlators with the expectations from a lossy SMSV GBS is characterized by a better agreement between data and model, as also quantified by the fitted value of the slope, equal to $\sim 0.94$ and thus close to the ideal value of $1$.

\begin{figure}[ht!]
    \centering
    \includegraphics[width=0.6\columnwidth]{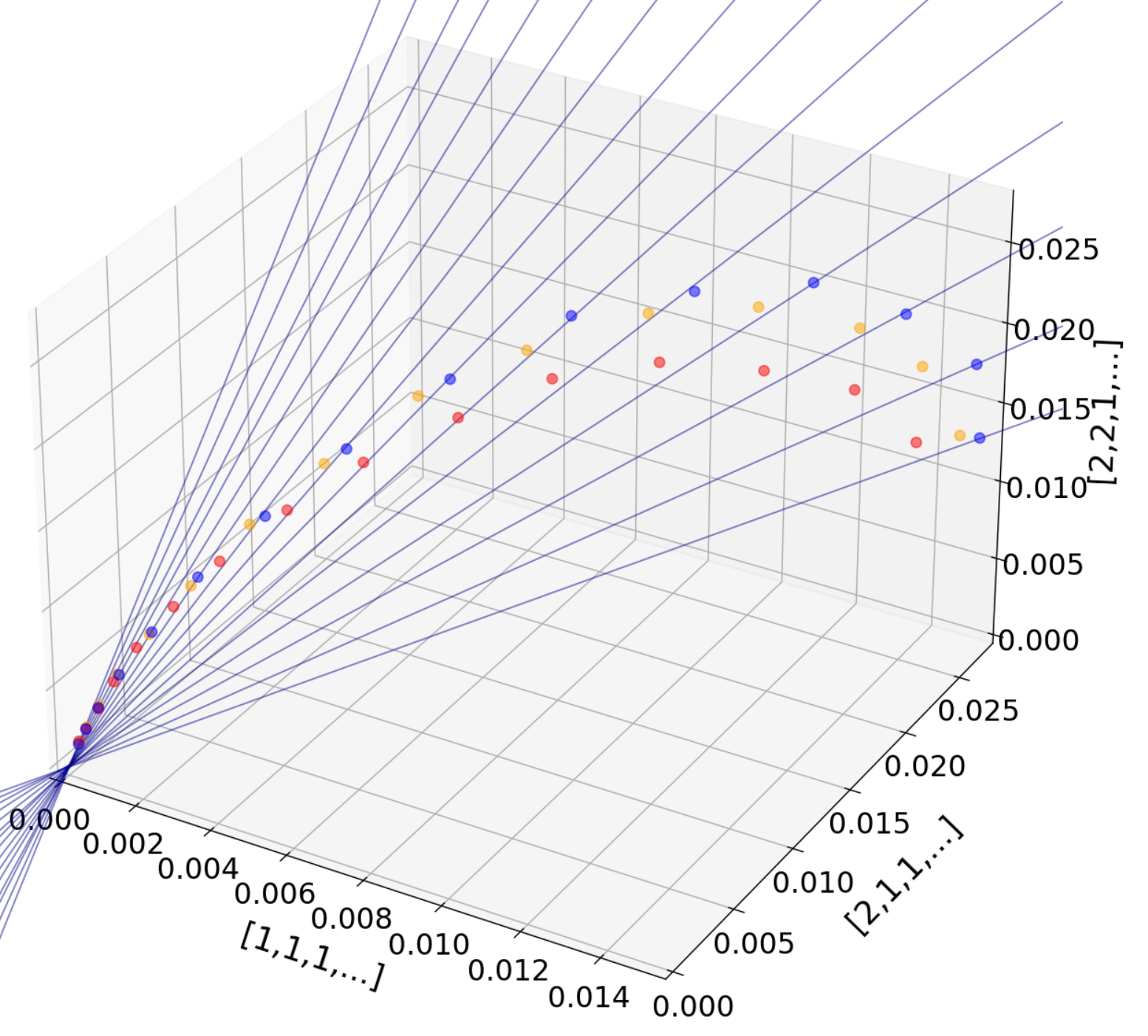}
    \caption{\textbf{Orbits of the data from the Borealis experiment of Ref. \cite{XanaduQAdv}.} In red Borealis data, in blue and orange the simulated data from the thermal and squashed sampler respectively. The size of the sample was $\sim 10^6$. The points correspond to the number of detected photons from 18 (from the right) to 32 (to the left). The Borealis data are those of Fig. 3a of Ref. \cite{XanaduQAdv}. The lines are those we obtained from the thermal simulations in Fig. \ref{fig:OrbitsThermal}.}
    \label{fig:borealis_data}
\end{figure}

These final observations allow us to conclude that, among the hypotheses tested in the above analysis, Borealis is likely to be performing a genuine SMSV GBS with losses. For completeness, we also analyzed some of the data from Ref. \cite{XanaduQAdv}, by using the method based on graph feature vectors. The results are shown in Fig. \ref{fig:borealis_data}, where we report in red the experimental orbits for the data corresponding to the lowest squeezing value of Ref.\cite{XanaduQAdv}. The simulated thermal (blue) and squashed (orange) samples were generated according to the circuit parameters of that time. It is worth noting that both the squeezing values and losses are smaller than the latest run with Borealis shown above. The orbits of the three models seem to be slightly more separated in this set of experimental data, in accordance with the expectation of a GBS device with reduced losses. The device certificates for this case can be found in the Supplemental Materials.

\section{Conclusions}
\label{sec:Conclusions}
In this work, we have analyzed the fundamental issue of assessing the operation of noisy intermediate scale quantum devices. More specifically, we have focused on the problem of Gaussian Boson Sampling in a noisy scenario, that represents one of the most investigated approaches to demonstrate the achievement of the quantum computational advantage regime. Here, we analyzed the capability of a graph theory based method to identify and exclude the main noise source in a Gaussian Boson Sampler. As the test platform, we have assessed the operation of Borealis, a large scale photonic device that has been made available on the cloud. In particular, we analyzed its working principles and on its performance as a sampling machine of indistinguishable SMSV states.

To this end, we first compared the feature vector components of the graph encoded in the device inspired by Ref. \cite{TairaGBS}, namely the orbits of Borealis, with those of the lossy thermal sampler and of a lossless SMSV GBS. This method is effective even in the quantum advantage regime as long as the number of detected photons is much smaller than the number of the modes. In such a condition, satisfied in Borealis only for the ``low'' level of the squeezing parameters, we investigated the effects of photon losses in the orbits estimation. In particular, the main source of noise in Borealis resulted to be the amount of balanced losses. Their effect is to move the orbits on the same hyperplane individuated by the data of indistinguishable thermal, squeezed, and coherent light emitters. This allowed us to exclude with high confidence the presence of significant photon distinguishability in the device. The distribution on such a hyperplane of Borealis data shows significant differences also with the points generated by coherent light. The orbits of the simulated thermal and squashed light are still close to the Borealis data instead. The small deviations observed could be compatible with the discrepancy between the device certificate employed to set the parameters of the simulations and the actual experimental conditions. Additional analysis have been also reported regarding the greedy algorithm mockup, simulating samples having the correct marginals up to a chosen order. As a further benchmarking of the device, we evaluated the two-mode correlation functions \cite{GBSVal3, GBS_correlators} summarized in the covariance matrix of samples. This last analysis showed a significant deviation from a lossy thermal sampler and a less pronounced, although still present, deviation from the squashed sampler, and thus shows that, among the class of hypotheses tested by the employed approach, Borealis is likely to behave as a lossy SMSV GBS. 

On one hand, our analysis showed the effectiveness of the orbits method in noisy conditions and, in particular, its power in highlighting the effect of the various contribution of photon losses in GBS experiments. On the other hand, the observation that the orbits generated by data collected from Borealis and by thermal and squashed samplers underline possible limitations in the use of the device for graphs-related problems. More precisely, our results showed that the collected feature vector components can be approximated by a classical simulation with thermal light, and by using squashed light the simulation can be even closer to the ones obtained from Borealis. Other recent works are opening questions regarding the advantage of the use of a noisy GBS device against simulation with classical simulatable models for the problem of graph max-clique and densest subgraphs \cite{GraphGBSNoAdvantage1, GraphGBSNoAdvantage2}. Due to the reduced connectivity of the circuits and to the limited range of the phase shifters achievable via the Borealis device, we were not able to find a way to encode a meaningful graph in the device, namely an adjacency matrix with a submatrix of significant size that is completely non-negative (or non-positive), to test the machine in this context. Furthermore, the structure of the device does not permit to encode graphs belonging to different classes of isomorphism, thus preventing the use of the graphs feature vectors, associated to the measured orbits, for graphs similarity and isomorphism applications. Future perspectives of our analysis regard a systematic study of noisy GBS devices such as Borealis for these graphs-related problems, and an investigation on whether the method for certification employed here can be extended to the original version of Boson Sampling, in light of recent studies of its connection with graph-theory\cite{Mezher23}.

\section*{Acknowledgements} 
We would like to thank Brajesh Gupt, Cedric Lin, and the Amazon Braket team at Amazon Web Services for the valuable discussion and support. This work is supported by the ERC Advanced Grant QU-BOSS (QUantum advantage via non-linear BOSon Sampling, Grant Agreement No. 884676) and by ICSC – Centro Nazionale di Ricerca in High Performance Computing, Big Data and Quantum Computing, funded by European Union – NextGenerationEU. D.S. acknowledges Thales Alenia Space Italia for supporting the Ph.D. fellowship. N.S. acknowledges funding from Sapienza Universit\`{a} di Roma via Bando Ricerca 2020: Progetti di Ricerca Piccoli, project n. RP120172B8A36B37.

\providecommand{\noopsort}[1]{}\providecommand{\singleletter}[1]{#1}%

\end{document}


\title{Supplemental materials for:Validation of a noisy Gaussian Boson Sampler via graph theory}

\author{Denis Stanev}
\affiliation{Gran Sasso Science Institute, Viale Francesco Crispi 7, I-67100 L’Aquila, Italy}
\author{Taira Giordani}
\affiliation{Dipartimento di Fisica, Sapienza Universit\`{a} di Roma, Piazzale Aldo Moro 5, I-00185 Roma, Italy}
\author{Nicol\`o Spagnolo}
\affiliation{Dipartimento di Fisica, Sapienza Universit\`{a} di Roma, Piazzale Aldo Moro 5, I-00185 Roma, Italy}
\email{nicolo.spagnolo@uniroma1.it}
\author{Fabio Sciarrino}
\affiliation{Dipartimento di Fisica, Sapienza Universit\`{a} di Roma, Piazzale Aldo Moro 5, I-00185 Roma, Italy}

\maketitle

\section{Analysis of the graphs method for Borealis validation in the lossless scenario}
The idea of the method is to arrange the output distributions of a gaussian boson sampler into the orbits. As described briefly in the main text, the orbits are the results of a coarse-graining of the output configurations. In this work, we considered the orbits of the first proposal in Ref. \cite{TairaGBS}. More precisely, given the number $n$ of detected photons, we have the orbit $O_1 = [1,1,1, \cdots, 0]$ that collects the output states in which the number of
photons in the modes is 0 or 1, $O_2=[2, 1,1, \cdots, 0]$  in which only one detector measures two photons, and $O_3=[2,2,1, \cdots, 0]$ in which two modes
host two photons. 
The various mock-up hypotheses differ for the \textit{feature vectors}, i.e. the vectors that have the probability of the orbits as components. Such vectors have a strong connection with graphs theory for an ideal gaussian boson sampler with indistinguishable squeezed vacuum states in the inputs.

The distribution of the vectors in the space spanned by only two or three orbit components displays different behaviors according to the nature of the gaussian boson sampler.
The method's theoretical proposal showed that collecting photon samples in the orbits from a small set of different unitary evolutions is enough to establish whether a classical model can emulate the experimental samples. The feature vectors extracted from the samples of a given model tend to cluster together by varying the unitary evolutions.  Furthermore, the estimate of the orbit probability can be obtained efficiently from a finite sample with the required accuracy for the validation task. This is largely studied and demonstrated in Ref. \cite{TairaGBS}.
Then, it is possible to establish how far are the feature vectors extracted from the single-photon samples from the mock-up hypotheses by looking, for example, at the distance between the centroids of the clusters or the dispersion and shape of the clusters. For clarity, we report
 a practical example to explain the method's functioning and effectiveness.
 
\begin{figure}
    \centering
    \includegraphics[width=0.5\linewidth]{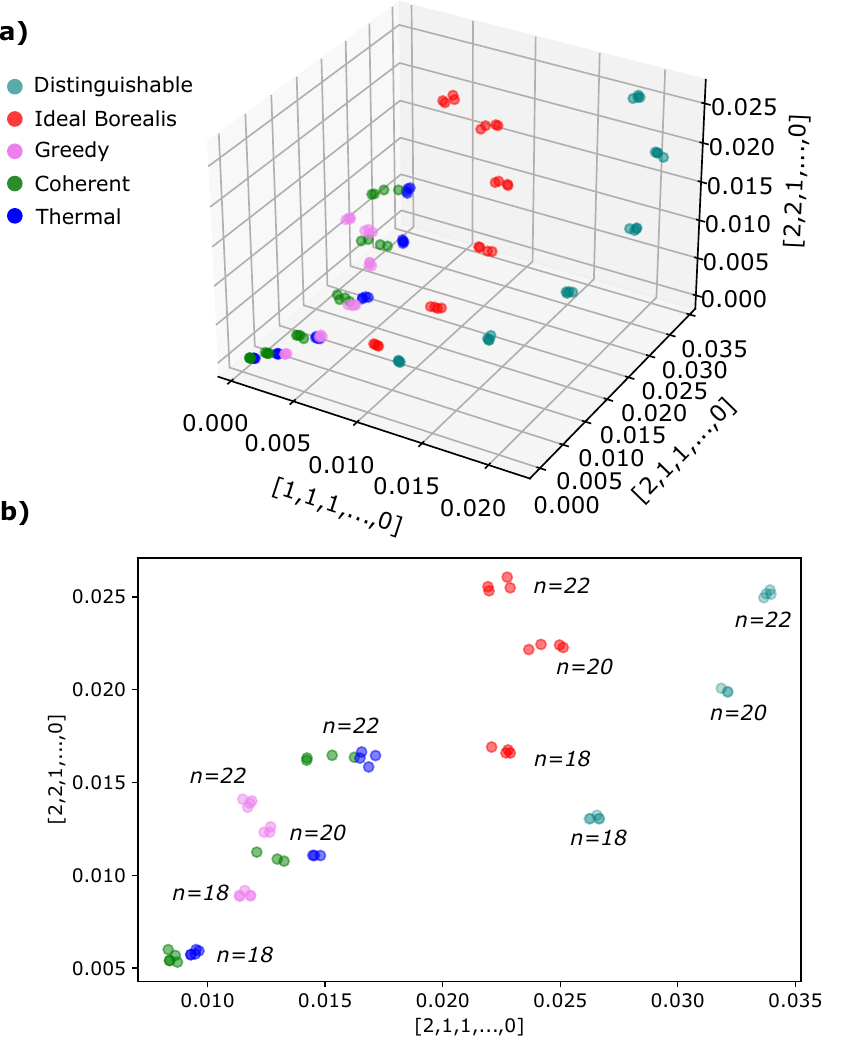}
    \caption{\textbf{Validation though the graphs method.} a) Distribution of the three orbits probabilities for the various mock-up hypotheses highlighted with different colors. The orbits of the mock-up hypotheses are retrieved from a sample of 250000 events generated by $m=216$ gaussian states sources that emit on average $n=26$ photons. The orbits correspond to 4 different configurations of the Borealis interfemoreter in a lossless condition. In red the ideal Borealis simulated data according to the method of Ref. \cite{TairaGBS}. b) Same analysis with a focus on the plane spanned by $O_2$ and $O_3$ for a number of post-selected photons equal to $n=18, \, 20, \, 22$.}
    \label{fig:example_orbits}
\end{figure}

In the simulation carried out in an ideal lossless condition, we consider the Borealis time-bin interferometer programmed to implement the 4 unitary transformations investigated in Fig.5b of the main text. Fig. \ref{fig:example_orbits} reports the distributions of the feature vectors of some mock-up hypotheses, like coherent and thermal light and distinguishable squeezed states. We also retrieved the feature vectors of a greedy sampler. All the data of the classical models are retrieved from sources that emit $n=26$ photons on average, on a sample of $250000$ events. 
 At this point, we estimate the orbits probability for an ideal device with indistinguishable squeezed vacuum states. In the regime of the simulation, it is still possible to approximate the orbits probability (and so the ground truth) by a Monte Carlo method used in the theoretical proposal \cite{TairaGBS}.
 Since such a simulation shows that the orbit probabilities of an ideal device are very well separated from the others, we can use the distance of the centroid of experimental feature vectors for a given set of unitary evolution from the ones of the classical models as an effective quantity for the validation task. Furthermore, we require that such a distance should be larger than the typical dispersion of the cluster.

\section{Simulation parameters and Borealis device certificate}
\label{sec:certificate}

Here we report a couple of the device certificates for Borealis. To be more precise, we report the device certificate for the 4th of April 2023, the day on which the measures for Fig. 5b were taken, and the device certificate for the 12th of January 2023, which is the base certificate for the thermal simulations of 4a. 

The device certificate shows the `loop\_phases' that refers to the 3 static phases $\tilde{\phi}_{1,2,3}$ that cannot be controlled by the user and represent the optical phases of each loop. Then we have the `squeezing\_parameters\_mean' which represents the squeezing values for the various presets. For most of our runs we employed the squeezer setting ``low''. The `common\_efficiency', `loop\_efficiencies', and `relative\_channel\_efficiencies' are the 3 types of efficiencies that we discuss in Section 4 of the main text. 

We use these parameters in the device certificate for the simulations. In the case of thermal, squashed, and coherent states simulations we generate the states so that in a lossless simulation the average number of photons is the same as that of the lossless simulation of the squeezed states. These values can be obtained directly from the squeezing value $s$. In the case of thermal and squashed states the mean photon-number population of the mode can be calculated as $\sinh^2(s)$ of the squeezing value, while for the coherent states, the displacement is calculated as $\sinh(s)$ of the squeezing value.

For the points of the thermal simulation in Fig. 4a, where we varied only the common efficiency, we started from the device certificate of the 12th of January, and we set the common efficiency to the values 0.350, 0.375, 0.400, 0.425 and 0.450, with the larger orbits corresponding to lower efficiencies. The green orbit, for which the detector number 5 was turned off to cause a significant unbalanced loss, was simulated with a common efficiency of 0.400. For the lossless indistinguishable SMSV states simulation we set the squeezing so that the average number of photons detected was 26. 

This is the device certificate for the 4th of April 2023, used for Fig. 5b:
\begin{lstlisting}[language=Python, showstringspaces=false]
{'finished_at': '2023-04-04
  T13:51:32.915878+00:00',
 'target': 'borealis',
 'loop_phases': [1.281, 0.671, 0.472],
 'schmidt_number': 1.135,
 'common_efficiency': 0.361,
 'loop_efficiencies': [0.875, 
 0.888, 0.717],
 'squeezing_parameters_mean': 
 {'low': 0.704, 
 'high': 1.178, 
 'medium': 0.998},
 'relative_channel_efficiencies': [0.925, 
  0.927, 0.915, 1.0, 0.965, 0.913,
  0.888, 0.968, 0.951, 0.946, 0.961,
  0.998, 0.952, 0.967, 0.946, 0.899]}
\end{lstlisting}

This is the device certificate for the 12th of January 2023, used for Fig. 4a:
\begin{lstlisting}[language=Python, showstringspaces=false]
{'finished_at': '2023-01-12
  T14:51:32.887242+00:00',
 'target': 'borealis',
 'loop_phases': [-0.797, 0.086, -2.689],
 'schmidt_number': 1.144,
 'common_efficiency': 0.392,
 'loop_efficiencies': [0.88, 
 0.836, 0.734],
 'squeezing_parameters_mean': 
 {'low': 0.669, 
 'high': 1.149, 
 'medium': 0.978},
 'relative_channel_efficiencies': [0.925,
  0.937, 0.914, 0.997, 0.978, 0.919,
  0.901, 0.971, 0.954, 0.956, 0.969,
  1.0, 0.942, 0.964, 0.96, 0.907]}
\end{lstlisting}

\section{Data from Borealis quantum advantage experiment}
\label{sec:borealis_data}

In the main text, we applied the method to some of the data from Ref. \cite{XanaduQAdv}.  In particular, the various efficiencies and squeezing value are the following:
\begin{lstlisting}[language=Python, showstringspaces=false]
{'loop_phases': [0.67, 3.239, 3.516],
 'common_efficiency': 0.475,
 'loop_efficiencies': [0.915, 
 0.895, 0.855],
 'squeezing_parameters_mean': 
 {'low': 0.533}
 'relative_channel_efficiencies': [0.978, 
 0.943, 0.958, 0.808, 0.924, 0.998, 
 0.893, 0.893, 0.985, 1.0, 0.888,
 0.936, 0.973, 0.921, 0.883, 0.951]}
\end{lstlisting}
while the beam-splitter splitting ratios and phase shifters values were set at random in the ranges [0.4, 0.6] and $[-\frac{\pi}{2}, \frac{\pi}{2}]$ respectively.

\providecommand{\noopsort}[1]{}\providecommand{\singleletter}[1]{#1}%